\documentclass[11pt,a4paper]{article}
\usepackage{jcappub}
\usepackage{amsmath,amssymb}
\usepackage{bm}
\usepackage{graphicx}
\usepackage{hyperref}
\hypersetup{colorlinks,bookmarksopen,bookmarksnumbered,citecolor=magenta,linkcolor=blue,pdfstartview=FitH,urlcolor=blue}

\usepackage{array}
\usepackage{physics}
\usepackage[mathscr]{eucal}
\usepackage{multirow}
\usepackage{xcolor}

\newcommand{\Mp}{M_{P}}
\renewcommand{\dd}{\mathrm{d}}
\newcommand{\cA}{\mathcal{A}}
\newcommand{\cB}{\mathcal{B}}
\newcommand{\cC}{\mathcal{C}}
\newcommand{\cD}{\mathcal{D}}
\newcommand{\cE}{\mathcal{E}}
\newcommand{\cH}{\mathcal{H}}
\newcommand{\cN}{\mathcal{N}}

\newcommand{\cL}{\mathcal{L}}
\newcommand{\adj}{\mathrm{adj}}

\title{\centering Can $n_s$ hint at a parity-odd\\ sector of gravity?}

\author[a]{Christian Dioguardi}
\emailAdd{christian.dioguardi@kbfi.ee}
\affiliation[a]{Laboratory of High Energy and Computational Physics, National Institute of Chemical Physics and Biophysics, R\"avala pst. 10, Tallinn, 10143, Estonia}

\abstract{We study slow-roll inflation in the context of metric-affine gravity. We consider the most general projectively invariant metric-affine theory at most quadratic in the distortion and linear in curvature with no propagating torsion and non-metricity and couple it non-minimally to a scalar field. We focus on the class of $\xi$-attractors, in which a single function $f(\phi)$ fixes the whole scalar sector. Integrating out the non-dynamical distortion generates a correction to the kinetic term $k(\phi)$ which is $f'(\phi)^2$ times a ratio of polynomials in $f(\phi)$. The new kinetic term can develop poles at finite field values, where the distortion field equations degenerate. We classify the poles and derive the conditions on the effective couplings for poles of order $p=2$ and $p=3$, the maximal order allowed in this setting. We then proceed to study inflation. Second-order poles realize $\alpha$-attractors with $\alpha$ fixed by the residue and are already available in the parity-even sector of the theory. Third-order poles instead require the parity-odd sector: whenever the parity-odd quadratic couplings vanish, the kinetic function splits into two independent terms with poles of order at most $p=2$. Third-order pole-inflation, as we confirm numerically for $f(\phi)\propto\phi^2$, yields $r\propto N_*^{-3/2}$ with a tunable amplitude and $n_s\simeq 1-\tfrac{3}{2N_*}$, a spectral index mildly larger than the $\alpha$-attractor prediction and well centered in the region preferred by the current combination of ACT, SPT and DESI data. Moreover, when inflation happens far away from the pole we also find a further universal attractor, an inverse-linear plateau, with $n_s \simeq 1-\frac{4}{3N_*}$ and $r\propto N_*^{-4/3}$, testable with the forthcoming BICEP/Keck observations.}

\begin{document}
\maketitle
\flushbottom

\section{Introduction}
\label{sec:intro}

Observations of the Cosmic Microwave Background (CMB) show that the universe is extremely homogeneous and flat \cite{Planck:2018vyg}. In order to explain such fine-tuned properties the Standard Cosmological Model, a.k.a. the $\Lambda$CDM, is supplemented by a period of quasi-exponential accelerated expansion that drives the universe to the observed flatness and homogeneity \cite{Guth:1980zm,Linde:1981mu,Starobinsky:1980te}. This period goes under the name of inflation and has the appealing property of seeding the current observed large-scale structure of the universe. In order to sustain inflation a new particle, the inflaton, is usually introduced to drive the early accelerated expansion. In the most simple realization this particle is usually taken to be a scalar field, embedded in General Relativity (GR). \\
However, in this setting, minimal models fail to predict the CMB observables, in particular the tensor-to-scalar ratio, which is currently bounded at $r\lesssim 0.034$ at 95\% confidence level by the combined observations of Planck, BICEP/Keck, ACT and SPT \cite{Planck:2018jri,BICEP:2021xfz,AtacamaCosmologyTelescope:2025nti,Balkenhol:2025wms}.
Several inflation models have been studied in different contexts trying to address such a suppressed tensor-to-scalar ratio \cite{Boubekeur:2005zm, Conlon:2005jm, Cicoli:2008gp,Ellis:2013xoa, Giudice:2014toa, Bauer:2008zj, Racioppi:2018zoy, Antoniadis:2018ywb,Enckell:2018hmo, Tenkanen:2020dge, Dioguardi:2021fmr,Dioguardi:2022oqu,Dioguardi:2023jwa,Bostan:2025zdt,Dioguardi:2025mpp,Dioguardi:2025vci, Dioguardi:2026dfw}. In particular Starobinsky \cite{Starobinsky:1980te} and Higgs \cite{Bezrukov:2007ep} inflation predict a value $r\sim 0.003$, $n_s \sim 0.967$ well inside the current bounds and measurable within the next decade CMB experiments \cite{SimonsObservatory:2018koc,LiteBIRD:2020khw}. On the other hand, recent measurements from DESI \cite{DESI:2024mwx,DESI:2025zgx}, combined with the ACT and SPT dataset \cite{AtacamaCosmologyTelescope:2025nti,SPT-3G:2025bzu, Balkenhol:2025wms}, have shifted the observed spectral index to $n_s = 0.9728 \pm 0.0029$ at $68\%$ confidence level, pushing the two most popular models over $2\sigma$ away from the central value\footnote{The current ACT+DESI analysis might be affected by a $\sim$ 2-3$\sigma$ tension in the BAO-CMB data \cite{Ferreira:2025lrd}, so it is currently impossible to say whether the shift in $n_s$ is just an artifact of this tension. For this reason in this paper we consider two different datasets, the first including only the CMB data, the second including the combination of CMB and DESI measurements.}.
New mechanisms for inflation then might be necessary to explain the updated observations.\\
One of the most recently studied frameworks in this context is Metric-Affine Gravity (MAG) \cite{Hehl:1976kj,Hehl:1976my,Hehl:1994ue}.
In the metric-affine formalism the metric $g_{\mu\nu}$ and the affine-connection $\Gamma^{\sigma}{}_{\mu\nu}$ are treated as \emph{a priori} independent objects (unlike in GR where the connection is assumed to be the Levi-Civita one). MAG contains in general several more degrees of freedom \cite{Baldazzi:2021kaf}, encoded in the two geometrical quantities that characterize a general connection: torsion $T^\sigma{}_{\mu\nu}$ and non-metricity $Q_{\sigma\mu\nu}$. Moreover in MAG, the so-called hypermomentum tensor naturally arises as a consequence of taking the variation of the action with respect to the independent connection \cite{Hehl:1976kt,Hehl:1978cb}, analogously to how one gets the energy-momentum tensor from the variation of the action with respect to the metric. The energy-momentum tensor and the hypermomentum are then to be regarded as the sources of MAG field equations \cite{BeltranJimenez:2020sih, Rigouzzo:2023sbb}.\\
The enriched particle spectrum of MAG carries, in general, several ghost degrees of freedom \cite{BeltranJimenez:2019acz,Percacci:2020ddy, Mikura:2023ruz, Mikura:2024mji, Barker:2025xzd, Percacci:2025oxw}. To avoid instabilities, one can then assume that the torsion and non-metricity are non-dynamical. In this way $T^{\sigma}{}_{\mu\nu}$ and $Q_{\sigma\mu\nu}$ only appear algebraically and can be eliminated from the field equations.
When the inflaton field is non-minimally coupled to torsion and non-metricity, however, the hypermomentum is sourced directly by the scalar field gradient inducing a modified inflationary dynamics with respect to the standard formulation in General Relativity \cite{Rigouzzo:2022yan, Andrei:2026xad}, even though torsion and non-metricity are non-dynamical. Several cases have been studied in this context\footnote{Inflation has also been studied in well-behaved sub-cases in which torsion and non-metricity are dynamical, see for example \cite{Salvio:2022suk, Gialamas:2022xtt, DiMarco:2023ncs, Gialamas:2024iyu, He:2024wqv, He:2025bli, Karananas:2025xcv, Salvio:2025izr,Iosifidis:2025wrv, Katsoulas:2025srh}. A notable, testable case has been recently studied in \cite{Dioguardi:2026hfe}, where the so-called $\mathring{R}^2$ scalaron is constructed from a symmetry principle and drives parameter-free inflation in very good agreement with current observations.} \cite{Shimada:2018lnm,Shaposhnikov:2020gts,Shaposhnikov:2020frq,Mikura:2020qhc,Mikura:2021ldx,Racioppi:2024zva,Gialamas:2024jeb,Capozziello:2024lsz,Racioppi:2024pno,Gialamas:2024uar,Racioppi:2025igu,Dimopoulos:2026iwq,Kraiko:2026nas,Hassan:2026ddx,Andrei:2026gmu} and the case of inflation in general MAG has only been recently addressed in \cite{Andrei:2026zvx}.
\\
In this recent paper, single-field slow-roll inflation was studied in the general case of projectively invariant metric-affine theory containing all operators at most linear in the curvature and quadratic in the distortion, non-minimally coupled to a real scalar field $\phi$. There, it has been shown that integrating out the non-dynamical distortion modifies the Jordan-frame scalar kinetic term by a factor $1+k(\phi)$. For $\xi$-attractors, i.e.\ when all the non-minimal couplings are linear in a single positive monotonic function $f(\phi)$ and the potential is $V(\phi)=M^4f(\phi)^2$, $k(\phi)$ is $f'(\phi)^2$ times a ratio of polynomials in $f(\phi)$, and in the strong-coupling regime the model universally produces an $\alpha$-attractor \cite{Kallosh:2013hoa,Ferrara:2013rsa,Kallosh:2013yoa,Kallosh:2013tua,Galante:2014ifa,Carrasco:2015uma} with
\begin{equation}
\alpha = 1+\frac{\zeta}{6\xi}\,,
\end{equation}
independently of the choice of $f(\phi)$, where $\xi$ is the non-minimal coupling with the Ricci scalar and $\zeta$ collects the whole effect of the metric-affine couplings. That analysis was restricted to the case of $k(\phi)$ regular at every finite field value, while its pole structure was left unexplored. In this paper we fill that gap.

In order to do so, we adopt the same $\xi$-attractor parametrization of the couplings as in \cite{Andrei:2026zvx}, keeping $f(\phi)$ generic. Our results can be summarized as follows:
(i) the order of poles at finite field values is bounded by $p\leq 3$, a direct consequence of the fact that a scalar gradient excites exactly three independent distortion modes;
(ii) second-order poles arise whenever the denominator of $k(\phi)$ has a double zero, and are already available in the parity-even sector of the theory: they realize $\alpha$-attractors in which $\alpha$ is a free parameter fixed by the residue at the pole;
(iii) third-order poles require the denominator to be a perfect cube, and, our main result, they are impossible unless the theory contains parity-odd operators: if these are vanishing, the kinetic function splits into two additive pieces whose denominators are respectively quadratic and linear in $f(\phi)$, so that no pole can exceed second order;
(iv) the pole-inflation predictions do not depend on the choice of $f(\phi)$, which only enters at subleading order.
Finally, we scan the parameter space to visualize the predictions of second-order and third-order pole inflation in the $(n_s,r)$ plane. For the numerical analysis we use, as a test case, the minimal choice $f(\phi)=\phi^2/\Mp^2$, i.e.\ a quartic Jordan-frame potential, while all the attractors we identify are derived for a generic $f(\phi)$.

The paper is organized as follows. In section \ref{sec:setup} we briefly recall the model and the outcome of integrating out the distortion. In section \ref{sec:kinetic} we present the Einstein-frame action, organize the pole structure of the modified kinetic term and show that pole inflation does not depend on the choice of $f(\phi)$. In section \ref{sec:poles} we derive the conditions for second- and third-order poles and the corresponding inflationary predictions in the slow-roll approximation. In section \ref{sec:parity} we show that the parity-even sector alone is not sufficient to generate third-order poles. We furthermore show that adding a scalar-Holst coupling restores this possibility. In section \ref{sec:parameters} we scan the parameter space to understand the general predictions of MAG in presence of a pole in the kinetic term. In section \ref{sec:conclusions} we discuss the validity of the results and draw our conclusions. In appendix \ref{app:1} we define all the projectively invariant quantities that enter the action, while in appendix \ref{app:coeffs} we collect all the required coefficient definitions which correspond to those found in \cite{Andrei:2026zvx}.
\section{The projective-invariant model}
\label{sec:setup}

We work on a four-dimensional non-Riemannian manifold $(\mathcal{M},g_{\mu\nu},\Gamma^{\rho}{}_{\mu\nu})$ with signature $(-,+,+,+)$ and an independent affine connection, following the conventions of \cite{Andrei:2026zvx}. The affine connection is split as $\Gamma^{\rho}{}_{\mu\nu}=\mathring{\Gamma}^{\rho}{}_{\mu\nu}+N^{\rho}{}_{\mu\nu}$, with $\mathring{\Gamma}$ the Levi-Civita connection and $N^{\rho}{}_{\mu\nu}$ the distortion tensor, equivalently encoded in the torsion $T^{\rho}{}_{\mu\nu}$ and the non-metricity $Q_{\rho\mu\nu}$,
\begin{equation}\label{eq:TQdef}
  T^\rho{}_{\mu\nu}\equiv2\Gamma^\rho{}_{[\mu\nu]},\qquad
  Q_{\rho\mu\nu}\equiv\nabla_\rho g_{\mu\nu},
\end{equation}
and related to the distortion through
\begin{equation}
  N_{\rho\mu\nu}=\tfrac12\big(Q_{\rho\mu\nu}-Q_{\nu\rho\mu}-Q_{\mu\rho\nu}\big)
                +\tfrac12\big(T_{\rho\mu\nu}+T_{\nu\rho\mu}+T_{\mu\rho\nu}\big).
  \label{eq:distortion}
\end{equation}
As in \cite{Andrei:2026zvx} we impose projective invariance to reduce the number of independent operators. The invariance under
\begin{equation}
\Gamma^{\rho}{}_{\mu\nu}\;\to\;\Gamma^{\rho}{}_{\mu\nu}+\delta^{\rho}{}_{\nu}\,\theta_{\mu}
\label{eq:projective},
\end{equation}
with $\theta_\mu$ an arbitrary vector, selects eight out of eleven parity-even scalars $J_i$, five out of six parity-odd scalars $O_i$ and three out of four linear terms in the distortion present in the general MAG action \cite{Iosifidis:2021bad,Iosifidis:2021tvx,Andrei:2026zvx}. Thus, the most general projectively invariant action at most quadratic in the distortion and linear in the curvature, non-minimally coupled to a real scalar, reads
\begin{align}
  \mathcal{S}=\int\!\dd^4x\sqrt{-g}\;\Big[\;
  &\tfrac12 F(\phi)\,R(\Gamma)
  +\partial_\mu\phi\,\big(\beta'_1(\phi)W^\mu+\beta'_2(\phi)U^\mu+\beta'_3(\phi)\bar T^\mu\big)+
  \nonumber\\[2pt]
  &+\sum_{i=1}^{8}\gamma_i(\phi)\,J_i
  +\sum_{i=1}^{5}\omega_i(\phi)\,O_i
  \;-\tfrac12(\partial\phi)^2-V(\phi)\Big],
  \label{eq:action_2}
 \end{align}
with $J_i$,$O_i$,$W^\mu, U^\mu, \bar{T}^\mu$ manifestly projectively invariant quantities defined in the appendix \ref{app:1} (see \cite{Andrei:2026zvx} for the details). The terms linear in the distortion arise from the Levi-Civita total derivatives $-\beta_1(\phi)\mathring\nabla_\mu W^\mu-\beta_2(\phi)\mathring\nabla_\mu U^\mu-\beta_3(\phi)\mathring\nabla_\mu \bar T^\mu$ after integration by parts, hence the notation $\beta'_i(\phi)$ \cite{Andrei:2026zvx}. This action contains eighteen free functions of $\phi$ (including the scalar potential $V(\phi)$). The distortion appears only algebraically: its field equations are linear and sourced by $\partial_\mu\phi$ alone. It can be proven that the general solution is given by \cite{Rigouzzo:2022yan}:
\begin{align}\label{eq:vectoransatz_1}
T_{\rho\mu\nu}&=t_1(\phi)\big(g_{\rho\mu}\partial_\nu\phi-g_{\rho\nu}\partial_\mu\phi\big)+t_2(\phi)\,\epsilon_{\rho\mu\nu\sigma}\partial^{\sigma}\phi\,,\\
Q_{\rho\mu\nu}&=q_1(\phi)\,\partial_\rho\phi\, g_{\mu\nu}+q_2(\phi)\big(g_{\rho\mu}\partial_\nu\phi+g_{\rho\nu}\partial_\mu\phi\big)\,.
\label{eq:vectoransatz_2}
\end{align}
It was shown furthermore in \cite{Andrei:2026zvx} that projective-invariance reduces the independent modes to three: by defining  $v\equiv(\sigma,q_2,t_2)^{\text{T}}$, with $\sigma\equiv 2t_1-q_1$ and using \eqref{eq:vectoransatz_1},\eqref{eq:vectoransatz_2} we can rewrite the algebraic sector of the Lagrangian as:
\begin{equation}
\cL_{\text{alg}}\big|_{\text{ansatz}}=\Big[\,S^{\text{T}}v+\tfrac12\,v^{\text{T}}M\,v\,\Big](\partial\phi)^2\,,
\qquad M=\tfrac32\hat M\,,\quad S=\tfrac32\hat{S}\,,
\label{eq:Lalg}
\end{equation}
with
\begin{equation}
\hat M=\begin{pmatrix} -\cA & -\cB & -\cE\\ -\cB & \cC & -\cH \\ -\cE & -\cH & \cD \end{pmatrix},
\qquad
\hat S=\begin{pmatrix} S_1 \\[2pt] S_2 \\[2pt] S_3 \end{pmatrix}\equiv\begin{pmatrix} F'+\beta'_1 \\[2pt] F'-\tfrac12 \beta'_1+3\beta'_2 \\[2pt] 4\,\beta'_3 \end{pmatrix},
\label{eq:MS}
\end{equation}
where the six \emph{effective couplings} $\cA,\ldots,\cH$ are the combinations of $F,\gamma_i,\omega_i$ listed in appendix \ref{app:coeffs}. Note for later use that the parity-odd couplings enter $\hat M$ only through
\begin{equation}
\cE=-6\,\omega_4\,,\qquad \cH=6\,\omega_3+3\,\omega_4-18\,\omega_5\,.
\label{eq:EHdef}
\end{equation}
Solving the linear system $Mv+S=0$ and substituting back yields:
\begin{equation}
\cL\;\supset\;-\tfrac12\big(1+k(\phi)\big)(\partial\phi)^2\,,\qquad
k(\phi)=S^{\text{T}}M^{-1}S=-\frac{3}{2\hat\Delta}\,\hat S^{\text{T}}\adj\big(\hat M\big)\hat S\,,\qquad
\hat\Delta\equiv-\det\hat M\,.
\label{eq:kdef}
\end{equation}
In \cite{Andrei:2026zvx} the analysis was restricted to $\det \hat M\neq 0$ at every field value; here we are interested precisely in the case $\hat\Delta(\phi_0)=0$, where $k(\phi)$ develops a pole. To render the analysis concrete we adopt the $\xi$-attractor parametrization of \cite{Andrei:2026zvx}, in which a single dimensionless function $f(\phi)$ fixes all the couplings and the potential,
\begin{equation}
\begin{gathered}
F=\Mp^2\big(1+\xi f(\phi)\big)\,,\qquad \gamma_i=\Mp^2\,\bar\gamma_i\,f(\phi)\,,\qquad \omega_i=\Mp^2\,\bar\omega_i\,f(\phi)\,,\\[2pt]
\beta'_i=\Mp^2\,\bar\beta_i\,f'(\phi)\,,\qquad V=M^4f(\phi)^2\,,
\end{gathered}
\label{eq:parametrization}
\end{equation}
with $M$ a constant mass scale and $f(\phi)>0$ a smooth, monotonically increasing and unbounded function, $f'(\phi)>0$\footnote{We work, with no loss of generality, in the $\phi\geq 0$ branch. Strictly speaking  $f'(\phi)>0$ only needs to hold in the region where inflation happens.} (from now on we drop the explicit $\phi$ dependence of $f(\phi)$ to simplify the notation). Every coupling is therefore linear in $f$ and the potential is quadratic in $f$. The minimal choice $f=\phi^2/\Mp^2$, for which $\gamma_i=\bar\gamma_i\phi^2$, $\omega_i=\bar\omega_i\phi^2$, $\beta'_i=2\bar\beta_i\phi$ and $V=\lambda\phi^4$ with $\lambda\equiv M^4/\Mp^4$, corresponds to a truncation of the action up to operators of dimension $d\leq 4$, and will be used as a test case in the numerical analysis of section \ref{sec:parameters}. Under \eqref{eq:parametrization} we have:
\begin{equation}
\begin{gathered}
\cA=\Mp^2(1+\bar a f)\,,\qquad \cB=\Mp^2(1+\bar b f)\,,\qquad \cC=\Mp^2(2+\bar c f)\,,\\[2pt]
\cD=\Mp^2(1+\bar d f)\,,\qquad \cE=\Mp^2\,\bar e f\,,\qquad \cH=\Mp^2\,\bar h f\,,
\end{gathered}
\label{eq:calligraphic}
\end{equation}
and $\hat S=\Mp^2 f'(\phi)\,(\bar s_1,\bar s_2,\bar s_3)^{\text{T}}$, with $\bar a,\ldots,\bar h$ and $\bar s_{1,2,3}$ given in appendix \ref{app:coeffs}.

\section{Einstein-frame kinetic function and pole-inflation}
\label{sec:kinetic}
By performing a conformal transformation of the metric
\begin{equation}\label{eq:weyl}
g^{E}_{\mu\nu}=\frac{F(\phi)}{M_P^2}\,g_{\mu\nu}\,,
\end{equation}
we can rewrite the theory in the equivalent Einstein frame\footnote{We assume throughout the whole paper that $F(\phi)>0$ for any $\phi$, which implies $\xi>0$ for the parametrization choice \eqref{eq:parametrization}.}:
\begin{equation}
\mathcal{S}=\int\dd^4x\sqrt{-g_E}\,\Big(\tfrac12 \Mp^2 R(g_E)-\tfrac12 K(\phi)(\partial\phi)^2-U(\phi)\Big)\,,
\end{equation}
with
\begin{equation}
K(\phi)=\Mp^2\left(\frac{1+k(\phi)}{F(\phi)}+\frac{3}{2}\frac{F'(\phi)^2}{F(\phi)^2}\right)\,,
\qquad
U(\phi)=\Mp^4\,\frac{V(\phi)}{F(\phi)^2}\,.
\label{eq:Kdef}
\end{equation}
By defining the canonical Einstein-frame scalar field as
\begin{equation}\label{eq:canonical_scalar}
\qty(\frac{\partial \chi}{\partial \phi})^2 = K(\phi)\,,
\end{equation}
we finally get the action in terms of the canonical scalar field:
\begin{equation}
\mathcal{S}=\int\dd^4x\sqrt{-g_E}\,\Big(\tfrac12 \Mp^2 R(g_E)-\tfrac12 (\partial\chi)^2-U(\chi)\Big)\,.
\end{equation}
We can now introduce the usual machinery needed to compute the slow-roll observables. We first define the slow-roll parameters:
\begin{equation}\label{eq:sr}
    \epsilon_U(\chi) \equiv \frac{M_P^2}{2}\qty(\frac{U'(\chi)}{U(\chi)})^2 ,\qquad \eta_U (\chi)\equiv M_P^2\frac{U''(\chi)}{U(\chi)},
\end{equation}
where the $'$ denotes the derivative with respect to its argument. The slow-roll parameters can be directly related to the CMB observables, in particular the tensor-to-scalar ratio $r$, the spectral index $n_s$, the amplitude of the scalar perturbations $A_s$ and the number of $e$-folds $N_*$. The latter does in general depend on the history of the evolution of the universe and hence, on the reheating temperature. In this paper we do not assume any specific reheating mechanism and consider the interval $50<N_*<60$. The aforementioned CMB observables can be expressed in terms of the slow-roll parameters $\epsilon_U, \eta_U$ as
\begin{align}\label{eq:sr_1}
r &= 16 \,\epsilon_U(\chi_*), \\ \label{eq:sr_2}
n_s &= 1 - 6\, \epsilon_U(\chi_*)+ 2 \,\eta_U (\chi_*),\\ \label{eq:sr_3}
A_s &= \frac{U(\chi_*)}{24\pi^2 M_P^4\,\epsilon_U(\chi_*)},\\ \label{eq:sr_4}
N_* &= \frac{1}{M_P^2} \int_{\chi_{\text{end}}}^{\chi_*}\frac{U(\chi)}{U'(\chi)} d\chi,
\end{align}
where $\chi_*$ is the canonical field at horizon crossing, and $\chi_{\text{end}}$ is set by $\epsilon_U(\chi) = 1$\footnote{Slow-roll can also end if $\eta_U(\chi) = 1$ is reached first, however, in most cases (such as the ones considered in this work) $\epsilon_U (\chi) = 1$ is reached first.}.
In the $\xi$-attractor class the potential is fixed by \eqref{eq:parametrization}, $V=M^4f^2$, so that $U=M^4f^2/(1+\xi f)^2$.
In general a simple analytic expression of
\eqref{eq:sr_1}-\eqref{eq:sr_4}, in terms of $\chi$ cannot be found, unless \eqref{eq:canonical_scalar} admits a closed form expression once integrated. However, this is not necessary: since $f$ is monotonic it can be used as field variable \cite{Andrei:2026zvx}, with
\begin{equation}
K_f(f)\equiv\qty(\frac{\partial\chi}{\partial f})^2=\frac{K(\phi)}{f'(\phi)^2}
=\frac{1+k(\phi)}{f'(\phi)^2\,(1+\xi f)}+\frac32\frac{\Mp^2\,\xi^2}{(1+\xi f)^2}\,,
\label{eq:Kf}
\end{equation}
and we can express \eqref{eq:sr_1}-\eqref{eq:sr_4} in terms of $f$ using the chain rule for derivatives:
\begin{align}
    r&=\frac{32\, M_P^2}{ K_f(f_*)\,f_*^2\left(1+\xi f_*\right)^2}\,,\\
    n_s&=1-\frac{2 M_P^2\left(4 K_f(f_*)+f_*  K_f'(f_* )\right)}{f_* ^2 \left(1+\xi f_*\right) K_f(f_*)^2}\,,\\
    A_s&=\frac{M^4  f_* ^4\, K_f(f_*)}{48 \pi ^2 M_P^6}\,,\\
    N_*&=\frac{1}{2M_P^2}\int_{f_{\text{end}}}^{f_*} f  \left(1+\xi f\right) K_f(f)\, df\,,
\end{align}
where $f_*\equiv f(\phi_*)$ is the value of $f(\phi)$ at horizon crossing and $K_f'\equiv\dd K_f/\dd f$. For $f=\phi^2/\Mp^2$ these reduce to the expressions in terms of $\phi_*$ for a quartic potential $V=\lambda\phi^4$.
Under the parametrization \eqref{eq:parametrization} the metric-affine modification to the kinetic term takes the form
\begin{equation}
k(\phi)=\Mp^2 f'(\phi)^2\,\tilde k(f)\,,\qquad
\tilde k(f)\equiv\frac{\cN(f)}{4\,\Delta(f)}
=\frac{\kappa_1+\kappa_2 f+\kappa_3 f^2}{4\big(3+ d_1 f+ d_2 f^2+d_3 f^3\big)}\,,
\label{eq:krational}
\end{equation}
with $\cN(f)\equiv\kappa_1+\kappa_2f+\kappa_3f^2$, $\Delta\equiv\hat\Delta/\Mp^6$ and the dimensionless coefficients $\kappa_{1,2,3}$ and $d_{1,2,3}$ collected in appendix \ref{app:coeffs}. For the minimal choice $f=\phi^2/\Mp^2$ eq.~\eqref{eq:krational} reads
\begin{equation}
k(\phi)=\frac{\Mp^4\,\kappa_1\phi^2+\Mp^2\,\kappa_2\phi^4+\kappa_3\phi^6}
{3\Mp^6+\Mp^4\, d_1\phi^2+\Mp^2\, d_2\phi^4+d_3\phi^6}\,.
\label{eq:krational_phi2}
\end{equation}
The denominator $\Delta(f)$ is a cubic polynomial in $f$: this degree is not accidental but counts the three distortion modes $(\sigma,q_2,t_2)$ reached by a scalar gradient, whose $3\times3$ structure in \eqref{eq:MS}, with entries linear in $f$, cannot generate anything higher. A source exciting more modes, or couplings of higher degree in $f$, could still give a correspondingly richer pole structure. Inserting \eqref{eq:krational} into \eqref{eq:Kf},
\begin{equation}
K_f(f)=\frac{\Mp^2\,\tilde k(f)}{1+\xi f}+\frac32\frac{\Mp^2\,\xi^2}{(1+\xi f)^2}+\frac{\Mp^2\,\alpha_1(f,f')}{f\,(1+\xi f)}\,,\qquad
\alpha_1(f,f')\equiv\frac{f}{\Mp^2f'^2}\,,
\label{eq:Kf_split}
\end{equation}
as in \cite{Andrei:2026zvx}: the first two terms depend on $f$ alone, while the functional form of $f(\phi)$ enters only through $\alpha_1(f,f')$ in the last term. For $f=\phi^2/\Mp^2$ one has $\alpha_1=1/4$, constant.

Since $F(\phi)>0$ is required for the conformal transformation \eqref{eq:weyl}, all finite-field poles of $K(\phi)$ descend from real zeros of $\hat\Delta$. The constant term $3\Mp^6$ never vanishes, so no pole can sit at $f=0$: poles sit at $\phi_0=f^{-1}(f_0)$, with $f_0>0$ a root of the cubic
\begin{equation}
\Delta(f)=3+ d_1\,f+ d_2\,f^2+d_3\,f^3\,.
\label{eq:cubic}
\end{equation}
Since $f'(\phi_0)\neq0$, a root of multiplicity $m$ in $f$ at $f_0$ is a zero of multiplicity $m$ in $\phi$ at $\phi_0$, so the pole order of $K(\phi)$ at $\phi_0$ is
\begin{equation}
p=m-n\,,
\end{equation}
with $n$ the multiplicity of $f_0$ as a zero of the numerator $\cN(f_0)$. Because \eqref{eq:cubic} is cubic,
\begin{equation}
p\;\leq\;3\,,
\end{equation}
and $p=3$ only if $\Delta$ is a perfect cube and $\cN(f_0)\neq0$. Together with the second-order pole at $f\to\infty$ studied in \cite{Andrei:2026zvx}, which yields the MAG $\alpha$-attractor, this exhausts the singularity structure of the kinetic function.

Since the Einstein-frame potential $U$ is regular in $f$, it is convenient to analyse the poles in the variable $f$. Near a pole the kinetic function behaves as
\begin{equation}
K_f(f)\simeq\frac{a_p}{(f_0-f)^p}\,,\qquad
a_p=\lim_{f\to f_0}(f_0-f)^p\,K_f(f)\,,
\end{equation}
while $U(f)$ is regular at $f_0$, with $U(f_0)>0$ and $U_{,f}(f_0)>0$. Expanding
\begin{equation}
    U\simeq U_0\big(1-\varepsilon_0\,(f_0-f)\big)\,, \qquad \varepsilon_0 = \frac{U_{,f}(f_0)}{U(f_0)}=\frac{2}{f_0(1+\xi f_0)}\,,
\end{equation}
and canonically normalizing, the standard pole-inflation results \cite{Galante:2014ifa} apply: for $p\geq2$ the pole is at infinite distance in the canonical Einstein frame field, the potential develops a plateau, which at large $e$-fold number predicts
\begin{equation}
n_s=1-\frac{p}{(p-1)N_*}\,,\qquad
r=\frac{8}{\big[(p-1)N_*\big]^{\frac{p}{p-1}}}\left(\frac{\varepsilon_0^{\,p-2}\,a_p}{\Mp^2}\right)^{\frac{1}{p-1}}\,.
\label{eq:polepredictions}
\end{equation}
The same expressions hold in any other field variable in which the potential is regular at the pole, e.g.\ $\phi$: under $\phi\to f(\phi)$ one has $a_p\to a_p\,f'(\phi_0)^{p-2}$ and $\varepsilon_0\to\varepsilon_0/f'(\phi_0)$, so the combination $\varepsilon_0^{\,p-2}a_p$, and hence $(n_s,r)$, is invariant. In the variable $f$, the pole term of \eqref{eq:Kf_split} depends only on $\tilde k(f)$, i.e.\ on $f_0$ and on the couplings, while the last term is regular at $f_0$. \emph{Pole-inflation predictions are therefore independent of the choice of $f(\phi)$.} 

The functional form of $f(\phi)$ first appears at order $(f_0-f)^0$ in the expansion of $K_f$ around the pole, through $\alpha_1(f,f')$, so that near the pole its effect is suppressed by $(f_0-f_*)^p$. In practice its leading effect comes from the last $e$-folds, where the field is far from the pole and $k(\phi)$ is no longer large, and amounts to an $\mathcal{O}(1)$ shift $\Delta N$ of the $e$-fold number, i.e.\ $\delta n_s\simeq -\tfrac{p}{p-1}\,\Delta N/N_*^2$, well within the uncertainty already associated with reheating\footnote{We checked this numerically for $f=(\phi/\Mp)^n$, $n=1,2,3$, and $f=e^{\phi/\Mp}-1$, with the benchmark couplings of section \ref{sec:flow}, finding $|\Delta N|\lesssim1$. The only requirement is $k(\phi_*)\gg1$, i.e.\ that at horizon exit the induced kinetic term dominates over the canonical one, which fails only if $f$ is extremely flat at the pole, $f'(\phi_0)\to0$.}. 

For $p=1$ the canonical field remains at finite distance, no plateau forms and at $f>f_0$ the kinetic term flips sign, implying instability; we do not discuss this case further. In the following we focus on the cases $p=2$ and $p=3$. Notice that odd-order poles need one extra assumption: $K_f(f)$ changes sign across $f_0$, so the pole is a boundary of field space and inflation must take place on the side where $K_f>0$.

\section{Conditions for second- and third-order poles}
\label{sec:poles}

\subsection{Second-order poles}
\label{sec:p2}

A second-order pole requires a double root of \eqref{eq:cubic} at $f_0>0$ with $\cN(f_0)\neq0$. Writing the discriminant of the cubic, the condition on the couplings is
\begin{equation}
d_1^2 d_2^2+54\, d_1 d_2 d_3-4\,d_1^3 d_3-12\, d_2^3-243\, d_3^2=0\,,
\qquad
\big(d_2^2,\;d_1 d_2\big)\neq\big(3\,d_1 d_3,\;27\, d_3\big)\,,
\label{eq:p2condition}
\end{equation}
which is a single relation among the $d_i$ (codimension one in coupling space) and the inequality excludes the perfect-cube configuration of section \ref{sec:p3}. The double root sits at
\begin{equation}
f_0=\frac{27\,d_3-d_1 d_2}{2\big(d_2^2-3\,d_1 d_3\big)}\,,
\label{eq:p2root}
\end{equation}
which must be positive, as well as the residue $a_2$, i.e.\ $\cN(f_0)>0$.
Close to the pole we have that:
\begin{equation}\label{eq:kpole_2}
    K_f(f) \simeq \frac{a_2}{(f_0-f)^2}\,, \qquad a_2 = \frac{\Mp^2\,\cN(f_0)}{2\,(1+\xi f_0)\,\Delta''(f_0)}\,,
\end{equation}
with $\Delta''(f_0)=2d_2+6d_3f_0$, where the residue of a second-order pole is the same in any field variable, $a_2=\lim_{\phi\to\phi_0}(\phi_0-\phi)^2K(\phi)$.
The resulting model is an $\alpha$-attractor,
\begin{equation}
n_s=1-\frac{2}{N_*}\,,\qquad r=\frac{12\,\alpha}{N_*^2}\,,\qquad
\alpha=\frac{2\,a_2}{3\Mp^2}=\frac{\cN(f_0)}{3\,(1+\xi f_0)\,\Delta''(f_0)}\,,
\label{eq:p2predictions}
\end{equation}
but of a different nature than the asymptotic one of \cite{Andrei:2026zvx}: there $\alpha$ was fixed by the large-field ratio $\zeta\equiv\kappa_3/d_3$ and the non-minimal coupling $\xi$, whereas here $\alpha$ is set by the residue of the pole at finite $f_0$. In particular, a small $\alpha$ (and hence small $r$) is achieved with a mild tuning of the residue rather than a large $\xi$ (as we also confirm numerically in section \ref{sec:parameters}).

It is worth stressing that second-order poles do \emph{not} require the parity-odd sector but do require the parity-even quadratic one. Setting $\cE=\cH=0$ the determinant factorizes, $\hat\Delta=\cD\big(\cA\cC+\cB^2\big)$, and the parity-even factor
\begin{equation}
\cA\cC+\cB^2=\Mp^4\big(3+P_1 f+P_2\,f^2\big)\,,\qquad
P_1=2\bar a+2\bar b+\bar c\,,\quad P_2=\bar a\bar c+\bar b^2\,,
\end{equation}
is quadratic in $f$ and supports a double zero under the simple conditions
\begin{equation}
P_1^2=12\,P_2\,,\qquad P_1<0\,,\qquad f_0=-\frac{6}{P_1}\,,
\label{eq:p2even}
\end{equation}
 and a non-vanishing numerator at $f_0$. Notice that a positive $f_0$ requires $P_1<0$. This cannot be achieved if the quadratic parity-even sector is turned off i.e. if $\gamma_i(\phi)=0$, in which case $P_1=6\xi>0$. Hence second-order poles are not available in absence of a quadratic parity-even sector (in the distortion). Equation \eqref{eq:p2even} is a one-parameter tuning of the parity-even quadratic couplings alone: $\alpha$-attractors are already available in parity-even projective-invariant metric-affine gravity.

\subsection{Third-order poles}
\label{sec:p3}

The maximal pole order $p=3$ requires the cubic \eqref{eq:cubic} to be a perfect cube. Since the constant term is fixed at $3$, this determines the whole polynomial in terms of $d_1$ alone:
\begin{equation}
d_2=\frac{d_1^2}{9}\,,\qquad d_3=\frac{d_1^3}{243}\,,\qquad d_1<0
\quad\Longrightarrow\quad
\Delta(f)=3\left(1-\frac{f}{f_0}\right)^{3},\qquad
f_0=-\frac{9}{d_1}\,.
\label{eq:p3condition}
\end{equation}
The inequality $d_1<0$ places the pole at a physical field value; via appendix \ref{app:coeffs}, $d_1=2\bar a+2\bar b+\bar c+3\bar d$, so a negative net quadratic coupling of the distortion sector is what pushes the degenerate zero $f_0$ onto the positive real axis. The pole survives in $k(\phi)$ provided the numerator does not vanish there,
\begin{equation}
\cN(f_0)\neq0
\quad\Longleftrightarrow\quad
\kappa_1\,d_1^2-9\,\kappa_2\, d_1+81\,\kappa_3\neq 0\,,
\label{eq:p3numerator}
\end{equation}
otherwise the order degrades. Conditions \eqref{eq:p3condition} define a codimension-two surface in the space of effective couplings; we comment on the associated tuning in section \ref{sec:flow}.

Near the pole, \eqref{eq:p3condition} gives $\Delta\simeq 3(f_0-f)^3/f_0^3$ and hence
\begin{equation}
K_f(f)\;\simeq\;\frac{a_3}{(f_0-f)^3}\,,\qquad
a_3=\frac{\Mp^2\,f_0^3\;\cN(f_0)}{12\,(1+\xi f_0)}\,,
\label{eq:kpole_3}
\end{equation}
so that a healthy kinetic term approaching $f\to f_0^-$ requires $\cN(f_0)>0$. The predictions follow from \eqref{eq:polepredictions} with $p=3$:
\begin{equation}
n_s=1-\frac{3}{2N_*}\,,\qquad
r=\frac{2\sqrt2}{N_*^{3/2}}\,\sqrt{\frac{\varepsilon_0\;a_3}{\Mp^2}}=\frac{2}{\sqrt3}\,\frac{f_0\sqrt{\cN(f_0)}}{(1+\xi f_0)\,N_*^{3/2}}\,.
\label{eq:p3predictions}
\end{equation}
Two features should be noted. First, unlike the $p=2$ case the tensor-to-scalar ratio retains a dependence on the potential through $\varepsilon_0$: the amplitude of $r$ is a free parameter scaling as $N_*^{-3/2}$, intermediate between the $N_*^{-1}$ of monomial models and the $N_*^{-2}$ of $\alpha$-attractors. Second, the spectral index is shifted upward with respect to the $\alpha$-attractor value: at $N_*=50$--$60$ one finds $n_s\simeq0.970$--$0.975$, to be compared with $n_s \simeq 0.960$--$0.967$ for $p=2$. Given that current data combinations including DESI BAO pull the preferred $n_s$ towards larger values \cite{Balkenhol:2025wms}, third-order pole inflation lands naturally in the middle of the allowed region, making the two universality classes distinguishable with forthcoming CMB data.
\section{Third-order poles and the parity-odd sector}
\label{sec:parity}

We now come to the main structural result of this paper: the conditions \eqref{eq:p3condition} cannot be satisfied, within the parametrization \eqref{eq:parametrization} and for any choice of $f(\phi)$, with a surviving third-order pole unless the parity-odd quadratic sector is switched on. The precise statement is the following.
 If $\cE=\cH=0$, i.e.
\begin{equation}
\omega_4=0\qquad\text{and}\qquad \omega_3=3\,\omega_5\,,
\label{eq:parityevencondition}
\end{equation}
then the kinetic function $k(\phi)$ has poles of order at most $p=2$, for arbitrary values of all remaining couplings. In particular, setting $\omega_3=\omega_4=\omega_5=0$ is sufficient (although not necessary) to forbid the third-order pole.

When $\cE=\cH=0$ the matrix \eqref{eq:MS} becomes block diagonal and the solution \eqref{eq:kdef} splits \emph{additively} into
\begin{equation}
k(\phi)\Big|_{\cE=\cH=0}
=\frac32\left[\frac{\cA\,S_2^2-2\,\cB\,S_1S_2-\cC\,S_1^2}{\cA\cC+\cB^2}
\;+\;\frac{S_3^{\,2}}{\cD}\right]\,,
\label{eq:ksplit}
\end{equation}
as it can be straightforwardly checked. Under the choice \eqref{eq:parametrization} the first denominator, $\cA\cC+\cB^2$, is \emph{quadratic} in $f$ and can have at most a double zero, while the second, $\cD$, is \emph{linear} in $f$ and can only have a simple zero at most. Since a sum of rational functions cannot have a pole of higher order than its addends, we have $p\leq2$.

Note that the conditions \eqref{eq:p3condition} can still be met with $\cE=\cH=0$ and they reduce to
\begin{equation}
P_1=6\,\bar d\,,\qquad P_2=3\,\bar d^2\,,
\label{eq:fakecube}
\end{equation}
i.e.\ the double zero of the parity-even factor is made to coincide with the zero of $\cD$, and $\hat\Delta$ indeed develops a triple zero at $f_0=-1/\bar d$. But when the two fractions in \eqref{eq:ksplit} are brought over the common denominator $\cD(\cA\cC+\cB^2)$, the numerator inherits a simple zero at $f_0$ from whichever factor multiplies each term, and the pole order remains $p=2$ at most. Note also that the parity-odd \emph{linear} coupling $\beta'_3(\phi)\,\partial_\mu\phi\,\bar T^\mu$ is of no help: it enters only the source $\hat S$, i.e.\ only the numerator of \eqref{eq:kdef}, and hence can never raise the order of the pole. We conclude that in order to have a third-order pole, a parity-odd term \emph{must} appear in the quadratic sector.

\subsection{From \texorpdfstring{$p=2$}{p=2} to \texorpdfstring{$p=3$}{p=3}}
\label{sec:flow}

An exact third-order pole lives on a codimension two subspace of the parameter space. It might then seem an unlikely configuration realized only in presence of fine-tuning of the parameters. Indeed, one must always keep in mind that a generic one-parameter deformation of the couplings does not even encounter $p=2$, since the double root already costs one condition, let alone $p=3$. The conditions \eqref{eq:p2condition} and \eqref{eq:p3condition} however, are not disjoint: the perfect-cube locus lies on the boundary of the double-root locus, and one can pass continuously from one to the other. It is then possible, \emph{within} the $\alpha$-attractor family \eqref{eq:p2condition}, to have  third-order points occur at isolated values of the remaining parameters, so a generic path along the $p=2$ surface can flow through a $p=3$ point.

Such a path is easy to construct explicitly. Keeping the double root fixed at $f_0$ and letting the third root $f_1$ vary,
\begin{equation}
\Delta(f)=3\left(1-\frac{f}{f_0}\right)^{2}\left(1-\frac{f}{f_1}\right)\,,
\qquad f_1\equiv f_0\,(1+\tilde{\epsilon})\,.
\label{eq:family}
\end{equation}
This fixes $d_{1,2,3}$ as functions of $\tilde{\epsilon}$, which in turn can be solved for three of the couplings. Without loss of generality, we choose $(\bar c,\bar d,\bar h)$, holding $(\bar a,\bar b,\bar e)$ and the source $\hat S$ fixed. The family is regular: $\bar c,\bar d,\bar h$ vary smoothly with $\tilde{\epsilon}$ and reach finite critical values at $\tilde{\epsilon}=0$, where \eqref{eq:family} becomes the perfect cube \eqref{eq:p3condition}.

Note that two poles are present for $\tilde{\epsilon}>0$: a double pole at $\phi_0=f^{-1}(f_0)$ and a simple pole at $\phi_1=f^{-1}(f_1)$. Their ordering is of fundamental importance, as we don't want to spoil the kinetic term in the region where inflation happens, hence we need $\phi_1\geq\phi_0$, i.e.\ $f_1\geq f_0$.
Writing
\begin{equation}
x=\phi_0-\phi\,,\qquad  \delta=\phi_1-\phi_0\,, \label{eq:delta}
\end{equation}
the kinetic function near $\phi_0$ takes the form
\begin{equation}
\;\;K(\phi)\;\simeq\;\frac{a_3}{f'(\phi_0)\,x^{2}\,(x+\delta)}\,,
\label{eq:crossover}
\end{equation}
with $a_3$ given in \eqref{eq:kpole_3}, which interpolates between the two universality classes in the single variable $x/\delta$:
\begin{equation}
K\simeq \frac{a_3}{f'(\phi_0)\,x^{3}}\quad (x\gg\delta)\,,
\qquad\qquad
K\simeq \frac{a_2}{x^{2}}\quad (x\ll\delta)\,,
\end{equation}
with
\begin{equation}
a_2=\frac{\Mp^2\,\cN(f_0)\,f_0^2f_1}{12\,(1+\xi f_0)\,(f_1-f_0)}\;\xrightarrow[f_1\to f_0]{}\;\frac{a_3}{f'(\phi_0)\,\delta}\,.
\label{eq:tworegimes}
\end{equation}
The inflaton must approach $\phi_0$ from below, i.e.\ from the side on which the simple pole lies beyond $\phi_0$: on the far side $K(\phi)$ changes sign across $\phi_1$ and the region $\phi>\phi_1$ will inevitably carry a ghost. The third-order point is therefore reached continuously \emph{only} from the $\alpha$-attractor side, which is what makes the flow well defined.

\begin{figure}[t]
\centering
\includegraphics[width=\textwidth]{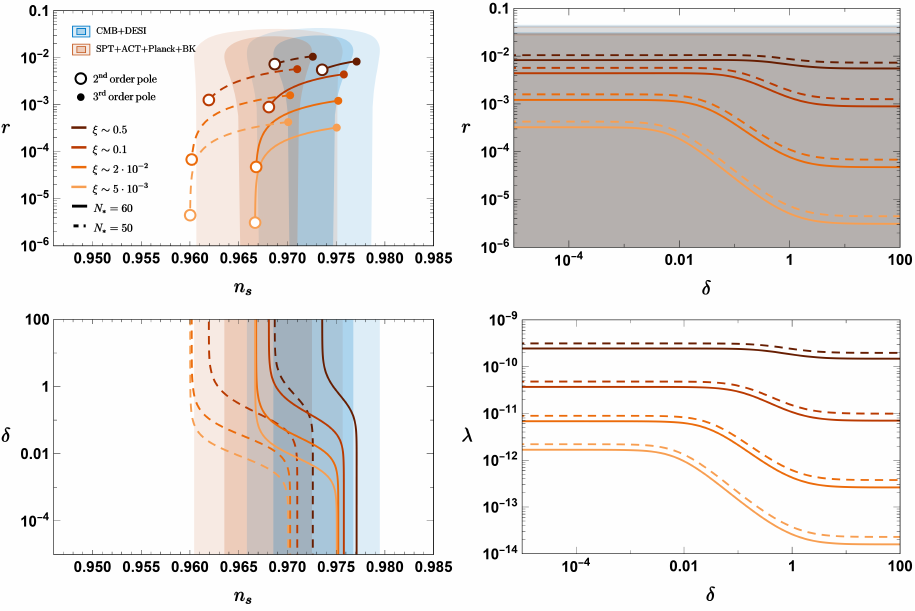}
\caption{Continuous flow from a second- to a third-order pole along the family \eqref{eq:family}, for $f=\phi^2/\Mp^2$, obtained by choosing $f_0=2.5$, $(\bar a,\bar b,\bar e)=(-0.55,-1.66,-0.80)$, for $\xi = 5\cdot 10^{-3}$ (lighter orange), $2\cdot 10^{-2}$(darker orange), $0.1$ (red), $0.5$ (brown) and setting $\bar\beta_{1,2,3}=0$, for simplicity. The observables $r, n_s$ were obtained by evaluating numerically the exact slow-roll equations \eqref{eq:sr_1}-\eqref{eq:sr_4} with the rational $K(\phi)$ of \eqref{eq:Kdef} for $V(\phi)=\lambda\phi^4$ Jordan frame potential, \emph{without} using any pole approximation. Panel (a) shows the trajectory in the $(n_s,r)$ plane as $\delta\rightarrow 0$ (with $\delta$ defined in \eqref{eq:delta}), with both $n_s$ and $r$ monotonically increasing. Panels (b) and (c) show that the entire flow is controlled by $\delta$, with the two plateaus separated by a crossover. Panel (d) shows the value of $\lambda$ fixed in terms of the other parameters through the measured amplitude of the scalar perturbation $A_s \sim 2.1\cdot 10^{-9}$. Details in the text.}
\label{fig:crossover}
\end{figure}

Figure \ref{fig:crossover} shows this flow for the family \eqref{eq:family} for the minimal choice $f=\phi^2/\Mp^2$. The specific example in the plot is obtained by choosing $f_0=2.5$ (i.e.\ $\phi_0^2=2.5\,\Mp^2$), $(\bar a,\bar b,\bar e)=(-0.55,-1.66,-0.80)$, for $\xi = 5\cdot 10^{-3}$ (lighter orange), $2\cdot 10^{-2}$(darker orange), $0.1$ (red), $0.5$ (brown) and setting $\bar\beta_{1,2,3}=0$, for simplicity. The observables $r, n_s$ were obtained by evaluating numerically the exact slow-roll equations with the rational $K(\phi)$ of \eqref{eq:krational_phi2} for $V(\phi)=\lambda\phi^4$ Jordan frame potential, \emph{without} using any pole approximation. Panel (a) shows the trajectory in the $(n_s,r)$ plane as $\delta\rightarrow 0$: it runs continuously from the second-order pole prediction to the third-order pole prediction, with $n_s$ and $r$ monotonically increasing. Panels (b) and (c) show that the entire flow is controlled by $\delta$, with the two plateaus separated by a crossover. Panel (d) shows the value of $\lambda$ fixed in terms of the other parameters through the measured amplitude of the scalar perturbation $A_s \sim 2.1\cdot 10^{-9}$.

We see that, in general, $\alpha$-attractor models are favored by the SPT+ACT+Planck+BK data, hence the current CMB-only observations show a preference for either second-order poles or the second-order pole at $f\to\infty$ studied in \cite{Andrei:2026zvx} (in the context of MAG inflation). These configurations can be already achieved by the model without including parity-odd terms in the Lagrangian.
However, with the inclusion of DESI data, observations seem to prefer either second-order poles with a large $\xi\gtrsim 0.5$\footnote{A large $\xi$ is sufficient but not necessary. As we are going to show numerically in the next section also alternative configurations with negligible $\xi$ can produce predictions compatible with the $(n_s,r)$ data from CMB+DESI.} or general third-order poles. The latter can only take place on a codimension two subspace of the full parameters space, but can also be reached continuously within the $\alpha$-attractor family by tuning the parameters. In either case, under parametrization \eqref{eq:parametrization} and for any choice of $f(\phi)$, third-order poles are a prediction of projective-invariant MAG inflation only in presence of a parity-odd gravitational sector and the current CMB+DESI dataset might represent an indirect hint at the presence of such non-minimally coupled operators.
We stress that the presence of such operators could only be confirmed by a direct measurement of parity-violation in the gravitational sector, a feature that cannot be confirmed by measurements of the spectral index alone.

\subsection{A minimal example: the Holst coupling}
\label{sec:holst}

The condition $(\cE,\cH)\neq(0,0)$ can be straightforwardly realized, for example, by a scalar-Holst non-minimal coupling $c_H(\phi)\tilde R(\Gamma)$, with $\tilde R(\Gamma)$ the contraction of the Riemann tensor with the Levi-Civita tensor:
\begin{equation}
\tilde{R} \equiv \epsilon^{\mu\nu\rho\sigma}R_{\mu\nu\rho\sigma},
\end{equation}
In the context of inflation, such a term has been extensively studied in several recent works \cite{Langvik:2020nrs, Racioppi:2024zva,Racioppi:2024pno,Dimopoulos:2026iwq}.
In our basis it corresponds to the choice $\omega_4=-\tfrac23 c_H$, $\omega_3-3\omega_5=c_H,\beta_3 = c_H$ \cite{Andrei:2026zvx}, which gives
\begin{equation}
\cE=\cH=4\,c_H\,,
\end{equation}
so a single Holst coupling switches on both parity-odd entries at once (for $c_H=\Mp^2\bar c_H f$, $\bar e=\bar h=4\bar c_H$). Third-order poles are then available and inflation with the Holst invariant thus emerges as one of the minimal projective-invariant settings in which the $p=3$ universality class \eqref{eq:p3predictions} can be realized (in presence of a non-vanishing parity-even sector).

\section{Parameter space and slow-roll inflation}\label{sec:parameters}

\begin{figure}[t]
\centering
\includegraphics[width=\textwidth,height=0.67\textheight,keepaspectratio]{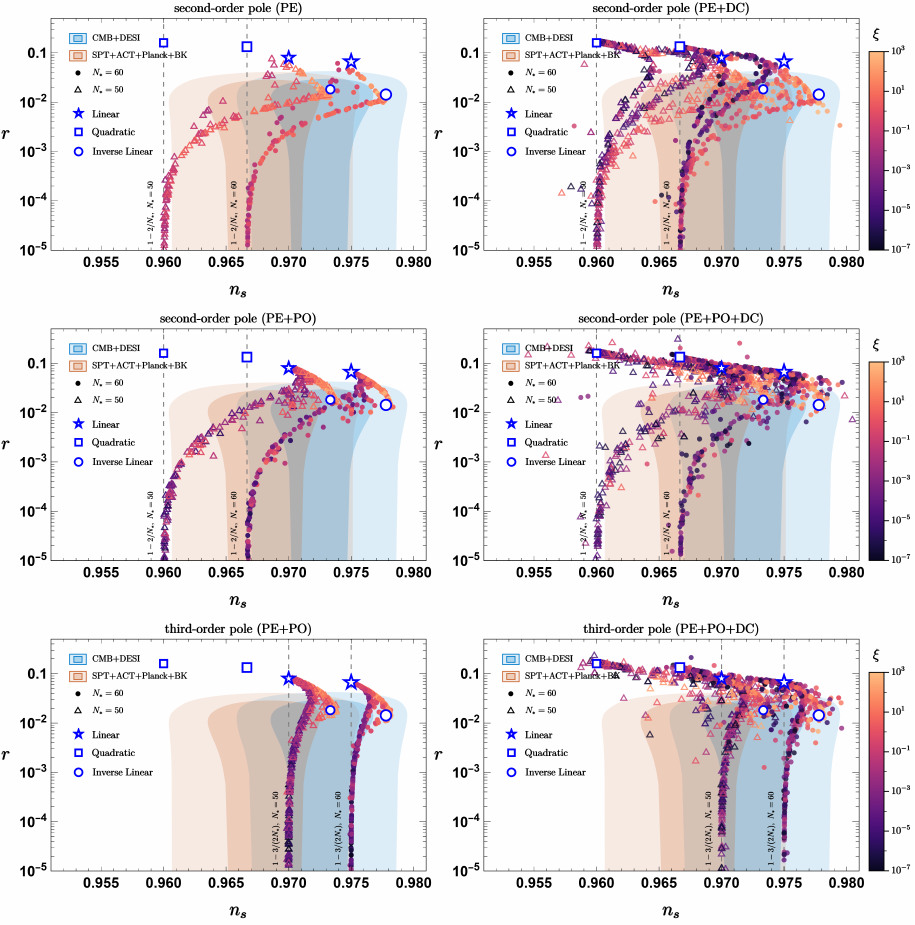}
\caption{Slow-roll predictions for the Jordan-frame potential $V(\phi) = \lambda \phi^4$, i.e.\ $f=\phi^2/\Mp^2$ in the parametrization \eqref{eq:parametrization}, for second- and third-order pole inflation with different sectors of the action \eqref{eq:action_2} switched on: a) (top-left) second-order pole, parity-even only (PE); b) (top-right) second-order pole, parity-even and linear distortion couplings (PE+DC); c) (center-left) second-order pole, parity-even and parity-odd (PE+PO); d) (center-right) second-order pole, all sectors (PE+PO+DC); e) (bottom-left) third-order pole, parity-even and parity-odd (PE+PO); f) (bottom-right) third-order pole, all sectors (PE+PO+DC). All observables are computed at $N_*=50$ (triangles) and at $N_*=60$ (bubbles); blue empty circles mark the inverse-linear plateau reached at large $\xi$ (section \ref{sec:largexi}, eq.~\eqref{eq:attractor} with $\bar a=\xi$), blue empty squares quadratic inflation (larger symbols for $N_*=60$). The orange contours show the 95\% and 68\% confidence levels based on the latest combination of Planck, BICEP/Keck, ACT and SPT \cite{Balkenhol:2025wms}, while the cyan contours show the 95\% and 68\% confidence levels after including DESI. Details in the text.}
\label{fig:scan}
\end{figure}

The projectively invariant action \eqref{eq:action_2} contains, under \eqref{eq:parametrization}, 17 independent couplings, $(\xi,\bar\gamma_{1,\dots,8},\bar\omega_{1,\dots,5},\bar\beta_{1,2,3})$. Only 14 of them are reached by the scalar gradient sourcing the distortion, since $\gamma_5$, $\omega_1$ and $\omega_2$ drop out of $\hat M$ and $\hat S$ (see appendix \ref{app:coeffs}). For the numerical scan we fix $f=\phi^2/\Mp^2$, i.e.\ $V=\lambda\phi^4$. We scan the parameter space by assuming $\xi>0$ and allow all the other couplings to be positive or negative. We sample the couplings in the range $10^{-7}$--$10^{3}$ (in absolute value) on a base-10 log scale, enforcing condition \eqref{eq:p2condition} for second-order and condition \eqref{eq:p3condition} for third-order poles, and compute $(n_s,r)$ for $\sim 10^{3}$ independent configurations by evaluating numerically \eqref{eq:sr_1}-\eqref{eq:sr_4}. The parameter scan assumes $|\phi_1| > |\phi_0|$ for second-order poles. Moreover, roughly $60-80\%$ of the sampled points (depending on the chosen subset) propagate a healthy kinetic term, while the remaining are ghosts\footnote{Unlike the predictions close to the pole, this fraction depends on the choice of $f(\phi)$, which enters the sign of $K_f$ away from the pole through the last, positive, term of \eqref{eq:Kf_split}.}. In Fig.\ref{fig:scan} we only plot the well-behaved configurations.
We show in Fig.\ref{fig:scan} the slow-roll predictions for the Jordan-frame potential $V(\phi) = \lambda \phi^4$ under the parametrization \eqref{eq:parametrization} with $f=\phi^2/\Mp^2$, for second- and third-order pole inflation with different sectors of the action \eqref{eq:action_2} switched on: a) (top-left) second-order pole, parity-even only (PE); b) (top-right) second-order pole, parity-even and linear distortion couplings (PE+DC); c) (center-left) second-order pole, parity-even and parity-odd (PE+PO); d) (center-right) second-order pole, all sectors (PE+PO+DC); e) (bottom-left) third-order pole, parity-even and parity-odd (PE+PO); f) (bottom-right) third-order pole, all sectors (PE+PO+DC). All observables are computed at $N_*=50$ (triangles) and at $N_*=60$ (bubbles); blue empty circles mark the inverse-linear plateau reached at large $\xi$ (section \ref{sec:largexi}, eq.~\eqref{eq:attractor} with $\bar a=\xi$), blue empty squares quadratic inflation (larger symbols for $N_*=60$). The orange contours show the 95\% and 68\% confidence levels based on the latest combination of Planck, BICEP/Keck, ACT and SPT \cite{Balkenhol:2025wms}, while the cyan contours show the 95\% and 68\% confidence levels after including DESI.
We stress that the points in Fig.~\ref{fig:scan} do not assume the pole approximation.

Besides the predictions \eqref{eq:p2predictions} and \eqref{eq:p3predictions}, obtained when $f_* \rightarrow f_0$, the scan displays limits in which
horizon exit occurs far away from the pole. They can all be understood in the same way, for a generic $f(\phi)$. For $f_*\ll f_0$, the denominator of $\tilde k(f)$ \eqref{eq:krational} becomes $\Delta\simeq 3$, so that $k(\phi)\simeq\Mp^2f'^2\,\cN(f)/12$, with $\cN$ a polynomial in $f$. If a single power dominates then
\begin{equation}
 k(\phi)\simeq\frac{\Mp^2 f'^2\,\kappa_n f^{n-1}}{12}\gg1\,, \qquad K_f(f)\simeq \frac{\Mp^2\,\kappa_n f^{n-1}}{12\,(1+\xi f)}\,,
\end{equation}
with $n=1,2,3$. In this limit we can integrate expressions \eqref{eq:sr_1}-\eqref{eq:sr_4} exactly (after neglecting the contribution coming from the end of inflation):
\begin{equation}
N_*\simeq\frac{\kappa_n\,f_*^{n+1}}{24(n+1)}\,,\quad
r\simeq\frac{16}{(n+1)(1+y_*)N_*}\,,\quad
n_s\simeq1-\frac{n+3+(n+2)\,y_*}{(n+1)(1+y_*)N_*}\,,
\label{eq:farpole}
\end{equation}
with
\begin{equation}
y_*\equiv\xi f_*\,.
\end{equation}
These limits do not depend on the choice of $f(\phi)$, which only affects their range of validity through the condition $k\gg1$.
For $y_*\ll1$ we have
\begin{equation}
 F\sim M_P^2\,,\qquad K_f \propto f^{n-1}\,, \qquad U\sim M^4 f^2\,, \qquad  \chi \propto f^{(n+1)/2}\,,
\end{equation}
which then implies
\begin{equation}
U(\chi)\propto\chi^{4/(n+1)},
\end{equation}
corresponding to quadratic inflation for $n=1$ and linear inflation for $n=3$.\\
For $y_*\gg1$ instead horizon exit takes place on the Jordan-frame plateau where
\begin{equation}
F\sim \Mp^2\xi f\,, \qquad K_f \propto f^{n-2}\,, \qquad
U\simeq U_0\qty(1-\frac{2}{\xi f})\,,\qquad \chi \propto f^{n/2}\,,
\end{equation}
which yields
\begin{equation}\label{eq:invplateau}
U(\chi) = U_0 \qty(1-\frac{U_1}{\xi\chi^{2/n}})\,,
\end{equation}
i.e.\ an inverse-power-law plateau approached at $\chi\to\infty$. Here the field rolls down from the plateau towards decreasing $\chi$, as in $\alpha$-attractors. We remind that, since we performed a random sampling, these attractor behaviors are to be considered as leading order approximations in specific regimes, so the approximation is never exact for the plotted points. In the next section we look closely at these attractors and understand which sector is responsible for each regime.

\subsection{Effect of parity-even and linear sectors}

First of all, we remark, as already mentioned in section \ref{sec:p2} that poles with $p=2$ can be reached only if the quadratic parity-even sector is on. From \eqref{eq:p2even} and \eqref{eq:p3condition} we see that a positive $f_0$ can only be obtained if we have respectively $P_1, d_1 < 0$ and from the coefficient definitions of appendix \ref{app:coeffs} we see that this cannot happen if $\bar\gamma_i = 0$ (this is why we show no plot without the parity-even sector). More generally, for $\bar\gamma_i=0$ one finds $\Delta=(1+\xi f)\big[3(1+\xi f)^2+Z f^2\big]$, whose zeros at $f>0$ are simple: the parity-odd sector alone can at most generate the unhealthy $p=1$ poles. Adding the linear sector does not allow to reach a higher pole order either: as discussed in section \ref{sec:parity}, the parity-even sector alone cannot generate third-order poles, and the $\beta'_i$ only enter the source $\hat S$, i.e.\ the numerator of $k(\phi)$.

The linear sector does, however, produce a new limit away from the pole: quadratic inflation. When the $\bar\beta_i$ dominate, so does the lowest power of $\cN(f)$\footnote{This can be understood by looking at the form of the coefficients $\kappa_{1,2,3}$ in appendix \ref{app:coeffs}. If $\bar\beta_i\gg \xi, \bar\gamma_i, \bar\omega_i$ then $\kappa_2$ and $\kappa_3$ are subdominant with respect to $\kappa_1$ that in this limit is determined by the $\bar\beta_i$ coefficients alone. Higher powers of $f$ could still be dominant if $f_*$ is large, however, in this regime $N_*\simeq\kappa_1 f_*^2/48$, so that $f_*\simeq(48N_*/\kappa_1)^{1/2}$ decreases as the $\bar\beta_i$ grow (for $f=\phi^2/\Mp^2$, $\phi_*$ is sub-Planckian \cite{Andrei:2026gmu}). Thus, only the lowest power $\kappa_1$ dominates in this case.},
\begin{equation}
k(\phi)\simeq\frac{\Mp^2\kappa_1 f'^2}{12}\,,  \qquad K_f\simeq\frac{\Mp^2\kappa_1}{12}\,,
\end{equation}
\begin{equation}
\kappa_1\simeq\tfrac92\big(12\bar\beta_2^2+64\bar\beta_3^2-12\bar\beta_1\bar\beta_2-\bar\beta_1^2\big)\qquad(\bar\beta_i\gg\xi)\,,
\end{equation}
which must be positive, i.e.\ driven by $\bar\beta_2$ or $\bar\beta_3$.
This corresponds to the case $n=1$, $y_*\ll1$ of \eqref{eq:farpole} i.e.
\begin{equation}
\chi=\sqrt{\frac{\kappa_1}{12}}\,\Mp f\,, \qquad U\simeq M^4f^2\propto\chi^2\,,
\end{equation}
for any $f(\phi)$, which predicts
\begin{equation}
n_s=1-2/N_*,\qquad r=8/N_*\,,
\end{equation}
in agreement with the results of \cite{Andrei:2026gmu,Andrei:2026zvx}. The quadratic limit is shown in Fig.\ref{fig:scan} with blue empty squares for $N_*= 50$ (smaller square) and $N_*=60$ (larger square) and only appears when the linear terms are switched on, as expected.

\subsection{Effect of the parity-odd sector}

If we add the parity-odd sector to the parity-even one, both second- and third-order poles are available, and both reach the corresponding plateau predictions \eqref{eq:p2predictions} and \eqref{eq:p3predictions}. Away from the pole a new limit appears: linear inflation. It corresponds to $n=3$, $y_*\ll1$ in \eqref{eq:farpole}:
\begin{equation}
K_f\simeq\frac{\Mp^2\kappa_3 f^2}{12}\,,\qquad  \chi\propto f^2,\qquad U\propto\chi\,,
\end{equation}
for any $f(\phi)$, which yields:
\begin{equation}
n_s=1-\frac{3}{2N_*}\,,\qquad r=\frac{4}{N_*}\,.
\end{equation}
This limit cannot be reached without parity-odd operators (see Fig.~\ref{fig:scan}, top-right). For $\cE=\cH=0$, eq.~\eqref{eq:ksplit} writes schematically as
\begin{equation}
k(\phi) \propto \Mp^2f'^2\left[\frac{\Mp^4\,f}{\cA\cC+\cB^2}+\frac{\Mp^2}{\cD}\right].
\end{equation} Far from the pole both denominators are constant, since $\hat\Delta=\cD(\cA\cC+\cB^2)\simeq3\Mp^6$, and $\tilde k(f)$ grows at most linearly in $f$. Therefore $n\leq2$ in the parity-even sector for $y_*\ll1$, and the limit of linear inflation, for which we need $n=3$, cannot be reached.

\subsection{All sectors and the large-\texorpdfstring{$\xi$}{xi} attractor}\label{sec:largexi}

With all sectors on we still find the plateau predictions of second- and third-order poles, together with the quadratic and linear limits, for different configurations of the parameters. Moreover, in all cases, for large $\xi$ and $\bar\beta_i\ll\xi$ (and in particular $\bar\beta_i=0$) the predictions converge to a new attractor, represented by the blue empty circles, corresponding to the inverse-power-law plateau \eqref{eq:invplateau} with $n=2$, i.e.\ an inverse-linear plateau $U\simeq U_0(1-\mu/\chi)$.\\

To show how this attractor is reached consider the following.
In this limit inflation takes place between the onset of the Jordan-frame plateau, $f\sim1/\xi$, and the pole at $f_0$. For large enough $f_0$ the window $1/\xi\ll f\ll f_0$ opens up at large $\xi$, and the value of the field at horizon exit falls inside it. There we still have $\Delta\simeq3$, while $F(\phi)\sim \Mp^2\xi f$. \\
Let us first consider the parity-even sector alone with $\bar\beta_i=0$.
Expanding $\tilde k$ at first order in $f$ one finds $\tilde k(f)\simeq\tfrac32\xi^2(\bar a f-1)$, independently of the pole condition \eqref{eq:p2even}, so that from \eqref{eq:Kf_split}, considering $f\ll f_0$, we get
\begin{equation}
K_f(f)\simeq\frac{3\Mp^2\xi^2 f\big[(\bar a-\xi)+\bar a\,\xi f\big]}{2(1+\xi f)^2}+\frac{\Mp^2\,\alpha_1}{f(1+\xi f)}\,,\qquad \bar a=\xi-3\bar\gamma_6\,.
\label{eq:Klargexi}
\end{equation}
Notice that this requires $\bar\gamma_6\leq 0$ in order to be well-defined everywhere. At large $\xi$ the first term dominates (for $f=\phi^2/\Mp^2$, $\alpha_1=1/4$, eq.~\eqref{eq:Klargexi} gives $K(\phi)\simeq\Mp^2/F+6\xi^2\phi^4[(\bar a-\xi)\Mp^2+\bar a\xi\phi^2]/(\Mp^2F^2)$). On the plateau, $\xi f\gg1$, one has
\begin{equation}
K_f\simeq \frac32\,\Mp^2\,\bar a\,\xi\,,\qquad\chi\propto f
\end{equation}
which gives
\begin{equation}
U\simeq U_0\left(1-\frac{2}{\xi f}\right)=U_0\left(1-\frac{\mu}{\chi}\right),\qquad \mu=\sqrt{\frac{6\bar a}{\xi}}\,\Mp\,,
\end{equation}
which is the inverse-linear plateau of \eqref{eq:invplateau}, up to corrections of order $1/(\xi f_*)$. This predicts:
\begin{equation}\label{eq:attractor}
n_s\simeq1-\frac{4}{3N_*}\,,\qquad r\simeq\frac{8}{(3N_*)^{4/3}}\left(\frac{6\bar a}{\xi}\right)^{1/3}\,.
\end{equation}
The couplings only enter through $(\bar a/\xi)^{1/3}$ in $r$. For $|\bar\gamma_6|\ll\xi$ one has $\bar a/\xi\to1$ and the attractor is completely universal and at leading order in $1/N_*$ we have:
\begin{equation}
n_s \simeq 1 - \frac{4}{3 N_*}\sim 0.9733,\qquad r\simeq \frac{8\cdot 6^{1/3}}{(3 N_*)^\frac{4}{3}}\sim 0.018\qquad \text{at $N_* = 50$}\,,
\end{equation}
which is testable and falsifiable within the next generation of CMB experiments \cite{SimonsObservatory:2018koc,LiteBIRD:2020khw}.
Note that this result is independent of the order of the pole and of the choice of $f(\phi)$, so both a second-order and third-order pole can reach this attractor. What is required is $\Delta\simeq3$ throughout the window $f\gg1/\xi$, i.e.\ a suppression of the coefficients $d_i$, which in our scan is realized by placing the pole at $f_0\gg1/\xi$.
The parity-odd sector adds to the expansion of $\tilde k$ a term $\tfrac32\xi^2\bar e^2f^2$, i.e.\ a contribution to $K_f$ growing linearly in $f$ on the plateau (as $\phi^4$ in $K(\phi)$ for $f=\phi^2/\Mp^2$). At horizon exit from our numerical scan we found it to be always subdominant with respect to $\tfrac32\Mp^2\bar a\xi$, leaving $n_s$ essentially unchanged (as one can see from Fig.\ref{fig:scan} in the center-left and bottom-left panels), unless the parity-odd couplings exceed $\xi$ by an order of magnitude or more, in which case the predictions drift toward linear inflation. Conversely, if the pole sits close to the onset of the plateau, $\xi f_0\lesssim1$, the $e$-folds are generated near the pole and one recovers the pole-inflation limit. This concludes the analysis of the parameter space.\\

Overall, we observe four distinct attractors: pole inflation at finite $f_0$, linear inflation, quadratic inflation, and the inverse-linear plateau, all of which are independent of the choice of $f(\phi)$. Any other point in the $(n_s, r)$ plane is either obtained from the competition of these four attractors or from ad-hoc fine-tuned choice of the couplings.
\section{Discussion and conclusions}
\label{sec:conclusions}

We studied single-field slow-roll inflation in the context of metric-affine gravity.
We considered the most general projectively invariant metric-affine gravitational sector up to linear order in curvature and quadratic order in the distortion, without propagating torsion and non-metricity. We coupled it with a real scalar field, through \emph{seventeen} independent functions $F(\phi), \gamma_i(\phi), \omega_i(\phi), \beta'_i(\phi)$. After integrating out the distortion and performing a conformal transformation we are left with an equivalent Einstein frame metric theory of gravity plus a propagating non-canonical scalar field and a modified scalar potential, so that all the contribution from the metric-affine gravitational sector is now encoded in the modified scalar field dynamics. Its kinetic term is determined by ten independent functions of the real scalar field $\mathcal{A},\mathcal{B},\mathcal{C},\mathcal{D},\mathcal{E},\mathcal{H}, F, S_1, S_2, S_3$ which are combinations of the original \emph{seventeen} free functions in the action. We focused on the class of $\xi$-attractors, in which every coupling is linear in a single monotonic function $f(\phi)$ and the potential is $V=M^4f^2$. In this class the kinetic term $K(\phi)$ is determined by $f(\phi)$ and by \emph{seven} effective parameters $\xi, \kappa_{1,2,3}, d_{1,2,3}$.
In a recent work, the case for which $K(\phi)$ has no poles was considered. There it was found that such a model always predicts an $\alpha$-attractor behavior at strong coupling, independently of $f(\phi)$ \cite{Andrei:2026zvx}.
Here we focused on the complementary case where $K(\phi)$ admits poles at finite field values. We characterized these poles and found that, within our assumption the model can only support finite-field poles of order $p\leq 3$, since the denominator of $k(\phi)$ is a cubic polynomial in $f$. In particular we found that poles of order $p=2$ are available only if at least one of the parity even sector coefficients $\bar\gamma_i \neq 0$, while $p=3$ only becomes available if in addition to that the parity-odd couplings switch on $(\cE,\cH)\neq(0,0)$, i.e.\ $\bar\omega_4\neq0$ or $\bar\omega_3\neq3\bar\omega_5$. We also found that $\bar\gamma_5, \bar\omega_{1,2}$ (or more generally $\gamma_5(\phi), \omega_{1,2}(\phi)$) do not affect the scalar field dynamics. Moreover the linear distortion couplings $\bar\beta_i$ have no effect on the pole order (but they still affect scalar field dynamics). We concluded that within the $\xi$-attractor class, for any choice of $f(\phi)$, second-order poles are already available in the parity even sector and third-order poles are realized only if we also include the parity-odd one. Moreover, the pole-inflation predictions do not depend on $f(\phi)$: they are fixed by the residue of the pole and by the slope of the potential at the pole, in a combination which is invariant under field redefinitions, while $f(\phi)$ only enters at subleading order through the canonical kinetic term of $\phi$, effectively shifting the number of $e$-folds by $\Delta N=\mathcal{O}(1)$.
For the minimal choice $f=\phi^2/\Mp^2$ we found that inflation happens close to the pole in the limit $\xi\ll1, \bar\beta_i\ll\xi$. In this limit the model predicts:
\begin{align}
    n_s &= 1-\frac{2}{N_*}\,,\qquad r = \frac{12\alpha}{N_*^2}\,,\qquad\hspace{46pt} \text{for\, second-order\, poles}\,,\\
    n_s &= 1-\frac{3}{2N_*}\,,\hspace{-6pt}\qquad r = \frac{2\sqrt2}{N_*^{3/2}}\,\sqrt{\frac{\varepsilon_0\;a_3}{\Mp^2}}\,\,,\qquad \text{for\, third-order\, poles}\,,
\end{align}
with $\alpha$ and $a_3$ given by \eqref{eq:p2predictions} and \eqref{eq:kpole_3}, and $\varepsilon_0=2/[f_0(1+\xi f_0)]$.
Comparison with data shows that with $50\leq N_* \leq 60$ second-order poles are preferred by the CMB-only data, while the CMB+BAO (i.e. including DESI data) prefers third-order poles. Since within the $\xi$-attractor class \eqref{eq:parametrization} third-order poles are only available in presence of a parity-odd sector we suggest that, if a larger $n_s$ value is selected by future observations, it could be a hint for the presence of a parity-odd sector of gravity, though, we remark, not conclusive. We furthermore underline that a third-order pole requires exact tuning. As proven in section \ref{sec:poles} a second-order pole costs one exact condition between the parameters \eqref{eq:p2condition} while a third-order one two \eqref{eq:p3condition}. As shown in section \ref{sec:flow} the second condition for the third-order pole can be relaxed and a significant $n_s$ shift towards third-order pole behavior can still be achieved within the $\alpha$-attractor family with tuning of the parameters, as explicitly shown in Fig.\ref{fig:crossover} for a benchmark example. The amount of tuning necessary for realizing such a scenario has only been shown qualitatively in this work and we postpone a more quantitative analysis for future work, in case larger values of $n_s$ are selected by future observations.

In section \ref{sec:parameters} we extended our analysis beyond pole-inflation and showed the general predictions for inflation, in presence of a pole in the kinetic term $K(\phi)$. We did so by performing a random scan of the parameter space for $f=\phi^2/\Mp^2$. Beyond the pole-inflation attractor we have shown, for a generic $f(\phi)$, that the model contains three more competing attractor behaviors, which are triggered by specific sectors of the gravitational action (see Fig. \ref{fig:scan}).
First of all, we found a general large $\xi$ attractor which appears whenever $\xi\gg 1 \gg \bar\beta_i$ and corresponds to an inverse-linear plateau which predicts, at leading order in $1/N_*$:
\begin{equation}
n_s \simeq 1 - \frac{4}{3 N_*}\sim 0.9733,\qquad r\simeq \frac{8\cdot 6^{1/3}}{(3 N_*)^\frac{4}{3}}\sim 0.018\qquad \text{at $N_* = 50$}\,,
\end{equation}
the attractor is available independently of the pole-order $p= 2,3$ and hence already available in the parity-even sector alone. Such a high value of the tensor-to-scalar ratio $r$ is in the observation range of Simons Observatory \cite{SimonsObservatory:2018koc} and BICEP/Keck \cite{BICEP:2021xfz} within the next few years and hence testable and falsifiable.
The model contains moreover two more attractors: linear inflation, available only when the parity-odd sector is present and quadratic inflation, which is triggered by the presence of linear distortion couplings as already shown in \cite{Andrei:2026gmu, Andrei:2026zvx} in related contexts. All these attractors are independent of the choice of $f(\phi)$, which only affects their range of validity.
Any other point in the $(n_s, r)$ plane generated by our random scan is either obtained as a result of the competition of these four attractors or as fine-tuned configurations of the parameter space.
In general we can conclude the following, depending on future observations of the tensor-to-scalar ratio.
If $r\gtrsim 10^{-3}$ our model can be compatible with data only outside the pole-inflation approximation. Hence if a pole is present, horizon exit in this model must happen when the scalar field is already far from $\phi_0$. In this regime, our model can predict specific testable values of the observables $n_s, r$ if $\xi\gg 1$ and $\bar\beta_i\ll \xi$.
On the other hand if $r\lesssim 10^{-3}$ our model can be compatible with data only if horizon exit happens close to $\phi_0$, which implies $\xi\ll 1,\bar\beta_i \ll \xi$. In this case the predictions of our model are specific and fall into the $\alpha$-attractor \eqref{eq:p2predictions} behavior if second-order pole inflation is realized, while they predict \eqref{eq:p3predictions} if third-order pole inflation is realized.
In particular, in our model third-order poles are the only configurations landing in the region preferred by the combination of CMB and BAO data for $r\lesssim 10^{-3}$: second-order poles give $n_s\simeq1-2/N_*$, while the quadratic and linear limits predict too large a tensor-to-scalar ratio. Third-order pole inflation thus can provide a mechanism to explain the observed shift of $n_s$ and, since in projectively invariant MAG such poles can only be obtained by switching on the parity-odd sector (within the $\xi$-attractor class, for any $f(\phi)$), recent observations might be hinting at the presence of such operators in the gravitational sector.

\acknowledgments
This work was supported by the Estonian Research Council via the grants, TARISTU24-TK10, TARISTU24-TK3, and the CoE program TK202 ``Foundations of the Universe''.

\appendix

\section{Projective invariants}\label{app:1}

We first define the projective invariants rank-3 tensors:
\begin{equation}
  \mathcal T_{\rho\mu\nu}=T_{\rho\mu\nu}+\tfrac13\big(g_{\rho\nu}T_\mu-g_{\rho\mu}T_\nu\big),
  \qquad
  \mathcal Q_{\rho\mu\nu}=Q_{\rho\mu\nu}-\tfrac14 g_{\mu\nu}Q_\rho ,
  \label{eq:traceless}
\end{equation}
from which we can construct two invariant parity-even vectors
\begin{equation}
  W_\mu=T_\mu-\tfrac38 Q_\mu,\qquad U_\mu=\tilde Q_\mu-\tfrac14 Q_\mu ,
  \label{eq:WU}
\end{equation}
together with a single parity-odd invariant vector:
\begin{equation}
  \bar T^\mu:=\epsilon^{\mu\nu\rho\sigma}T_{\nu\rho\sigma}
        =\epsilon^{\mu\nu\rho\sigma}\mathcal T_{\nu\rho\sigma},
  \label{eq:axial}
\end{equation}
with $\epsilon^{\mu\nu\rho\sigma}$ the Levi-Civita tensor and $T_\mu=T^\nu{}_{\nu\mu}$, $Q_\mu=Q_\mu{}^\nu{}_\nu$, $\tilde Q_\mu=Q^\nu{}_{\nu\mu}$ the trace vectors.
From these we can construct by contraction all the possible non-zero manifestly projectively invariant scalars (at most quadratic in the distortion). The parity-even ones read
\begin{alignat}{3}
  &J_1= \mathcal{T}_{\rho\mu\nu} \mathcal{T}^{\rho\mu\nu},\quad
  &&J_2= \mathcal{T}_{\rho\mu\nu} \mathcal{T}^{\mu\rho\nu},\quad
  &&J_3= \mathcal{Q}_{\rho\mu\nu} \mathcal{Q}^{\rho\mu\nu},\nonumber\\
  &J_4= \mathcal{Q}_{\rho\mu\nu} \mathcal{Q}^{\mu\rho\nu},\quad
  &&J_5= \mathcal{Q}_{\rho\mu\nu} \mathcal{T}^{\mu\rho\nu},\quad
  &&J_6=W_\mu W^\mu,\nonumber\\
  &J_7=W_\mu U^\mu,\quad
  &&J_8=U_\mu U^\mu, && \label{eq:Jbasis}
\end{alignat}
while the parity-odd terms are
\begin{alignat}{3}
  &O_1=\epsilon^{\mu\nu\rho\sigma} \mathcal{T}^\lambda{}_{\mu\nu} \mathcal{T}_{\lambda\rho\sigma},\quad
  &&O_2=\epsilon^{\mu\nu\rho\sigma} \mathcal{Q}_{\mu\rho\lambda} \mathcal{Q}_{\nu\sigma}{}^{\lambda},\quad
  &&O_3=\epsilon^{\mu\nu\rho\sigma} \mathcal{T}^\lambda{}_{\mu\nu} \mathcal{Q}_{\rho\sigma\lambda},\nonumber\\
  &O_4=W_\mu \bar T^\mu,\quad
  &&O_5=U_\mu \bar T^\mu . && \label{eq:Obasis}
\end{alignat}
We refer to \cite{Andrei:2026zvx} for the details.

\section{Coefficients}
\label{app:coeffs}

For the reader's convenience we collect from \cite{Andrei:2026zvx} the definitions used in the main text. The effective couplings entering \eqref{eq:MS} are
\begin{equation}
\begin{aligned}
\cA&=F-3\gamma_6\,,\\
\cB&=F+\tfrac32\gamma_6-\tfrac92\gamma_7\,,\\
\cC&=2F+12\gamma_3+3\gamma_4+\tfrac34\gamma_6-\tfrac92\gamma_7+27\gamma_8\,,\\
\cD&=F-8(\gamma_1-\gamma_2)\,,\\
\cE&=-6\,\omega_4\,,\\
\cH&=6\,\omega_3+3\,\omega_4-18\,\omega_5\,.
\end{aligned}
\label{eq:appAcouplings}
\end{equation}
Under the $\xi$-attractor parametrization \eqref{eq:parametrization} the effective couplings and the source take the form \eqref{eq:calligraphic}, $\hat S=\Mp^2f'(\bar s_1,\bar s_2,\bar s_3)^{\text {T}}$, with the dimensionless coefficients
\begin{equation}
\begin{aligned}
\bar a&=\xi-3\bar\gamma_6\,, &\qquad \bar s_1&=\xi+\bar\beta_1\,,\\
\bar b&=\xi+\tfrac32\bar\gamma_6-\tfrac92\bar\gamma_7\,, & \bar s_2&=\xi-\tfrac12\bar\beta_1+3\bar\beta_2\,,\\
\bar c&=2\xi+12\bar\gamma_3+3\bar\gamma_4+\tfrac34\bar\gamma_6-\tfrac92\bar\gamma_7+27\bar\gamma_8\,,\qquad & \bar s_3&=4\bar\beta_3\,,\\
\bar d&=\xi-8(\bar\gamma_1-\bar\gamma_2)\,,\\
\bar e&=-6\bar\omega_4\,,\\
\bar h&=6\bar\omega_3+3\bar\omega_4-18\bar\omega_5\,.
\end{aligned}
\end{equation}
With
\begin{equation}
P_1=2\bar a+2\bar b+\bar c\,,\qquad P_2=\bar a\bar c+\bar b^2\,,\qquad Z=2\bar e\bar h+2\bar e^2-\bar h^2\,,
\end{equation}
the numerator coefficients of \eqref{eq:krational} are
\begin{equation}
\begin{aligned}
\kappa_1&=6\Big[\bar s_2^2+3\bar s_3^2-2\bar s_1^2-2\bar s_1\bar s_2\Big]\,,\\
\kappa_2&=6\Big[\bar s_2^2(\bar a+\bar d)+\bar s_3^2P_1-\bar s_1^2(\bar c+2\bar d)-2\bar s_1\bar s_2(\bar b+\bar d)-2\bar s_1\bar s_3(\bar h+2\bar e)+2\bar s_2\bar s_3(\bar h-\bar e)\Big]\,,\\
\kappa_3&=6\Big[\bar s_2^2(\bar a\bar d+\bar e^2)+\bar s_3^2P_2-\bar s_1^2(\bar c\bar d-\bar h^2)-2\bar s_1\bar s_2(\bar b\bar d+\bar e\bar h)-2\bar s_1\bar s_3(\bar b\bar h+\bar c\bar e)+2\bar s_2\bar s_3(\bar a\bar h-\bar b\bar e)\Big]\,,
\end{aligned}
\label{eq:kappas}
\end{equation}
and the denominator coefficients are
\begin{equation}
d_1=P_1+3\bar d\,,\qquad
d_2=P_2+\bar d\,P_1+Z\,,\qquad
d_3=\bar d\,P_2+2\bar b\bar e\bar h+\bar c\bar e^2-\bar a\bar h^2\,,
\label{eq:ds}
\end{equation}
so that $\hat\Delta=\Mp^6\Delta(f)=\Mp^6\big(3+d_1f+d_2f^2+d_3f^3\big)$.
\bibliographystyle{JHEP}
\bibliography{references}

\end{document}